\documentclass[%
 aip,jcp,
 amsmath,amssymb,
 reprint,
]{revtex4-2}

\usepackage[a-1b]{pdfx} 

\usepackage{graphicx}
\usepackage{dcolumn}
\usepackage{bm}

\usepackage[utf8]{inputenc}
\usepackage[T1]{fontenc}
\usepackage{mathptmx}
\usepackage{listings}
\usepackage{rotating} 

\usepackage{color}
\usepackage{dcolumn} 
\usepackage{bm} 
\usepackage{graphicx}
\usepackage{multirow} 
\usepackage{pifont} 
\usepackage{epsfig}
\usepackage{amsmath} 
\usepackage{subfigure}
\usepackage{float}
\usepackage{booktabs}
\usepackage{tabularx}
\usepackage{natbib}
\usepackage{gensymb}
\usepackage{rotating}
\usepackage{threeparttable}
\usepackage{comment}
\usepackage[normalem]{ulem}

\begin{document}     

\author{Justin T. Phillips}
\affiliation{
             Department of Chemistry and Biochemistry,
             Florida State University,
             Tallahassee, FL 32306-4390}          

\author{Lauren N. Koulias}
\affiliation{
             Department of Chemistry and Biochemistry,
             Florida State University,
             Tallahassee, FL 32306-4390}  

\author{Stephen H. Yuwono}
\affiliation{
             Department of Chemistry and Biochemistry,
             Florida State University,
             Tallahassee, FL 32306-4390}

\author{A. Eugene DePrince III}
\email{adeprince@fsu.edu}
\affiliation{
             Department of Chemistry and Biochemistry,
             Florida State University,
             Tallahassee, FL 32306-4390}

\title{Comparing Perturbative and Commutator-Rank-Based Truncation Schemes in Unitary Coupled-Cluster Theory}

\begin{abstract}

Unitary coupled-cluster (UCC) theory offers a promising Hermitian alternative to conventional coupled-cluster (CC) theory, but its practical implementation is hindered by the non-truncating nature of the Baker-Campbell-Hausdorff (BCH) expansion of the similarity-transformed Hamiltonian ($\bar{H}$). To address this challenge, various truncation strategies have been developed to approximate $\bar{H}$ in a compact and reliable manner. In this work, we compare the numerical performance of approximate UCC with single and double excitations (UCCSD) methods that employ many-body perturbation theory (MBPT) and commutator-rank-based truncation schemes. Our results indicate low-order MBPT-based schemes, such as UCC(2) and UCC(3), yield reasonable results near equilibrium, but they become unreliable at stretched geometries. Higher-order MBPT-based schemes do not necessarily improve performance, as the UCCSD(4) and UCCSD(5) amplitude equations sometimes lack solutions. In contrast, commutator-rank-based truncations exhibit greater numerical stability, with the Bernoulli representation of the BCH expansion enabling more rapid convergence to the UCCSD limit compared to the standard BCH formulation.

\end{abstract}

\maketitle

\section{Introduction}

\label{SEC:INTRODUCTION}

Coupled-cluster (CC)\cite{Coester58_421,Kuemmel60_477,Cizek66_4256,Cizek69_35,Shavitt72_50,Li99_1,Musial07_291} theory is generally accepted as the gold standard for high-accuracy quantum chemistry calculations. This sterling reputation stems from the rapid convergence of truncated CC approaches [CC with singles and doubles (CCSD),\cite{Bartlett82_1910,Zerner82_4088} CC with singles, doubles, and triples (CCSDT),\cite{Bartlett87_7041,Schaefer88_382} CC with singles up to quadruples (CCSDTQ),\cite{Adamowicz91_6645,Bartlett91_387,Bartlett92_4282,Adamowicz94_5792} \emph{etc.}] to the exact, full configuration interaction (CI) limit. Moreover, only connected and linked terms arise in the CC energy and wave function expressions, respectively, which guarantees size extensivity and separability of the ground-state CC energy, even for truncated cluster operators. These nice properties notwithstanding, CC theory suffers from some well known problems that are related to the non-Hermiticity of the  the similarity-transformed Hamiltonian, $\bar{H}$. For example, a variety of numerical issues may arise when $\bar{H}$ is expanded within a truncated many-particle basis, such as complex energies\cite{Hattig05_37,Koch17_164105,Gauss21_e1968056,DePrince23_044113} and other unphysical properties\cite{DePrince23_044113} in the vicinity of conical intersections, as well as reduced density matrices that violate basic ensemble $N$-representability conditions.\cite{DePrince23_054113, Coleman63_668} 
As such, it is worth considering alternative {\em ans{\"a}tze} that retain CC theory's desirable properties, while eliminating the non-Hermiticity of the similarity-transformed Hamiltonian.

{Several alternatives to traditional CC theory have been proposed that aim for full Hermiticity or to reduce the degree of non-Hermiticity,} including the the expectation-value CC (XCC),\cite{Noga88_29,Trucks89_125} variational CC (VCC),\cite{Laforgue78_805,Cizek79_463,Noga88_29,Kutzelnigg91_349,Scuseria18_044107} and unitary CC (UCC)\cite{Kutzelnigg77_129,Koch83_4315,Kutzelnigg84_822,Simons88_993,Noga89_133,Bartlett95_281,Bartlett06_3393,Knowles10_234102,Evangelista11_224102,Scuseria18_044107,Mukherjee18_244110,Cheng21_174102,Cheng22_2281} approaches (see, e.g., Ref.~\citenum{Bartlett95_281} for a systematic assessment of these and other alternative CC {\em ans{\"a}tze}). None of these methods have supplanted traditional CC theory in practical calculations, though, because their working equations are characterized by infinite summations that must be artificially truncated in order to obtain manageable programmable expressions.  Focusing on UCC, the challenge is that the introduction of an anti-Hermitian cluster operator results in a non-truncating Baker--Campbell--Hausdorf (BCH) expansion of the similarity-transformed Hamiltonian, which contrasts with conventional CC approaches where the BCH expansion automatically truncates after four nested commutators. Two primary strategies are used to truncate $\bar{H}$, based on many-body perturbation theory (MBPT) analysis of the UCC amplitude and energy equations,\cite{Noga89_133,Bartlett89_359} or commutator rank.\cite{Simons88_993,Mukherjee18_244110,Cheng21_174102,Cheng22_2281} On the MBPT side, we have methods such as UCC(4),\cite{Noga89_133} which is a UCC approach with single, double, and triple excitations where the energy expression includes terms up to fourth-order in perturbation theory. 
On the other hand, the BCH expansion could simply be truncated at a predetermined commutator rank, without any perturbation theory considerations.\cite{Simons88_993} More recently, a compact commutator-rank based truncation scheme has been proposed based on the Bernoulli number expansion of the similarity-transformed Hamiltonian.\cite{Mukherjee18_244110,Cheng22_2281,Cheng21_174102} One notable feature of the Bernoulli expansion approach is that the Fock operator does not appear in any commutators of rank greater than one.

The goal of this paper is to examine the relative performance of MBPT- and commutator-rank-based truncation schemes in the context of calculations carried out at the UCC with single and double excitations (UCCSD) level of theory. To the best of our knowledge, such systematic studies are lacking in the literature, which is surprising, given that multiple numerical studies have examined the properties of these approaches in isolation. To that end, we assess the quality of electronic energies obtained from various approximations to UCCSD, as applied to a variety of small molecular systems at equilibrium and non-equilibrium geometries, as well as the classic Be + H$_2$ insertion reaction.\cite{Bartlett83_835,Evangelista11_224102}

The remainder of this paper is organized as follows. Section \ref{SEC:THEORY} provides an overview of CC and UCC theory and the truncation schemes used in this work.  Section \ref{SEC:COMPUTATIONAL_DETAILS} describes the relevant computational details. We discuss our findings for the small molecules and Be + H$_2$ insertion reaction in Section \ref{SEC:RESULTS}, and concluding remarks can be found in Section \ref{SEC:CONCLUSIONS}.

\section{Theory}

\label{SEC:THEORY}

In this section, we provide the relevant details of the CC and UCC formalisms and highlight the key components in the MBPT-based and Bernoulli representations of the BCH expansion. Throughout the discussion, the labels $i, j, \ldots$ (or $i_1, i_2, \ldots$) and $a, b, \ldots$ (or $a_1, a_2, \ldots$) refer to spin orbitals that are occupied and unoccupied in the reference configuration, respectively. The Einstein summation convention is used, where repeated lower and upper indices are summed.

The ground-state CC wave function takes the form
\begin{equation}
\label{eqn:cc_wfn}
    |\Psi_{\text{CC}}\rangle = e^{\hat{T}} |\Phi\rangle,
\end{equation}
where $|\Phi\rangle$ is a reference Hartree-Fock (HF) determinant and $\hat{T}$ is the cluster operator, which is defined as
\begin{equation}
\label{eqn:cc_co}
    \hat{T} = \sum_{n}^M \hat{T}_n,\; \hat{T}_n = \left(\frac{1}{n!}\right)^2 t_{a_1 \dots a_n}^{i_1 \dots i_n}
              \hat{a}^{a_1}\cdots\hat{a}^{a_n} \hat{a}_{i_n}\cdots\hat{a}_{i_1}.
\end{equation}
Here, $t_{a_1 \dots a_n}^{i_1 \dots i_n}$ is a $n$-body cluster amplitude, and $\hat{a}_p$ ($\hat{a}^p \equiv \hat{a}_p^{\dagger}$) is the fermionic annihilation (creation) operator acting on the spin-orbital labeled $p$. Specific methods within the hierarchy of truncated CC approaches are obtained by setting the truncation level, $M$ to a particular integer value, such as 2 for CCSD, 3 for CCSDT, etc., and the full CC method is obtained at $M = N$, where $N$ is the number of electrons.

Inserting the CC wave function into the Schr{\"o}dinger equation and left multiplying by $e^{-\hat{T}}$, we arrive at 
\begin{equation}
\label{eqn:cc_se}
    \bar{H} |\Phi\rangle = E |\Phi\rangle,
\end{equation}
where $\bar{H} = e^{-\hat{T}} \hat{H} e^{\hat{T}}$ is the similarity transformed Hamiltonian. The $\bar{H}$ operator can be expanded using the Baker--Campbell--Hausdorff (BCH) formula,
\begin{equation}
\label{eqn:cc_bch}
    \bar{H} = \hat{H} + [\hat{H}, \hat{T}] + \frac{1}{2!} [[\hat{H}, \hat{T}], \hat{T}] + \frac{1}{3!} [[[\hat{H}, \hat{T}], \hat{T}], \hat{T}] + \cdots,
\end{equation}
which naturally truncates after four nested commutators because the electronic Hamiltonian contains only one- and two-body interactions. The cluster amplitudes, $t_{a_1 \dots a_n}^{i_1 \dots i_n}$, are obtained by solving the projective equations
\begin{equation}
\label{eqn:cc_proj}
    \langle\Phi_{i_1 \cdots i_n}^{a_1 \cdots a_n}|\bar{H}|\Phi\rangle = 0 \quad \forall |\Phi_{i_1 \cdots i_n}^{a_1 \cdots a_n}\rangle, n=1,2,3, \ldots, M
\end{equation}
and the CC energy is then given by the expectation value
\begin{equation}
\label{eqn:cc_e}
    E = \langle\Phi|\bar{H}|\Phi\rangle.
\end{equation}

As mentioned in Section \ref{SEC:INTRODUCTION}, the CC similarity-transformed Hamiltonian is not Hermitian. The UCC formalism addresses this issue through the use of a similar exponential wave function {\em ansatz}
\begin{equation}
\label{eqn:ucc_wfn}
    |\Psi_{\text{UCC}}\rangle = e^{\hat{\sigma}} |\Phi\rangle,
\end{equation}
where $\hat{\sigma}$ is an anti-Hermitian cluster operator 
\begin{equation}
\label{eqn:anti_herm_cluster_operator}
    \hat{\sigma} = \hat{T} - \hat{T}^\dagger.
\end{equation}
As above, $\hat{\sigma}$ can be expressed as a many-body expansion with the components
\begin{align}
\label{eqn:sigma_mb_expansion}
    \hat{\sigma}_n & = \left(\frac{1}{n!}\right)^2  \Big[t_{a_1 \dots a_n}^{i_1 \dots i_n} \hat{a}^{a_1}\cdots\hat{a}^{a_n} \hat{a}_{i_n}\cdots\hat{a}_{i_1} \nonumber \\
                   & {} - t_{i_1 \dots i_n}^{a_1 \dots a_n *} \hat{a}^{i_1}\cdots\hat{a}^{i_n} \hat{a}_{a_n}\cdots\hat{a}_{a_1} \Big],
\end{align}
where the second term inside the square bracket represents the de-excitation part of $\hat{\sigma}$, which is not present in the conventional CC formalism. In this work, we use real-valued integrals, orbitals, and amplitudes, which means that $t_{i_1 \dots i_n}^{a_1 \dots a_n *} = t_{a_1 \dots a_n}^{i_1 \dots i_n}$. In analogy to Eqs.~\ref{eqn:cc_se}--\ref{eqn:cc_e}, the cluster amplitudes and UCC energy are obtained from the connected cluster form of Schr{\"o}dinger equation. However, the similarity-transformed Hamiltonian now takes the form $\bar{H} = e^{-\hat{\sigma}} \hat{H} e^{\hat{\sigma}}$, and, because $\hat{\sigma}$ is anti-Hermitian ({\em i.e.}, $\hat{\sigma}^\dagger = -\hat{\sigma}$), the exponentiated operator is unitary ({\em i.e.}, $e^{-\hat{\sigma}} = e^{\hat{\sigma}^\dagger}$). As a result, $\bar{H}$ maintains the Hermiticity of the original Hamiltonian. An unfortunate consequence of the use of the anti-Hermitian cluster operator is that BCH expansion,
\begin{equation}
\label{eqn:ucc_bch}
    \tilde{H} = \hat{H} + [\hat{H}, \hat{\sigma}] + \frac{1}{2!} [[\hat{H}, \hat{\sigma}], \hat{\sigma}] + \frac{1}{3!} [[[\hat{H}, \hat{\sigma}], \hat{\sigma}], \hat{\sigma}] + \cdots,
\end{equation}
no longer truncates, at any order. As a result, practical implementations of UCC methodologies can only be realized by forcing the truncation of the series. One challenge is that the choice of truncation scheme is not unique.

In the earliest UCC studies, the BCH expansion in Eq.~\ref{eqn:ucc_bch} was truncated using MBPT-based arguments.\cite{Noga89_133} Assuming a canonical HF reference, the Fock ($f$) and fluctuation ($v$) operators appear at the 0th and 1st orders in MBPT, respectively, and the  $\hat{T}_n$ operator appears at ($n-1$)th order (with the exception of $\hat{T}_1$, which appears at 2nd order). The UCC($n$) hierarchy of method restricts the BCH expansions in the energy {equation to contain up to $n$th order terms, and the residual equations are obtained by making the energy stationary with respect to variations in the amplitudes. As a result, the highest-order terms that appear in the residual equations are ($n-1$)th order for the doubles residual equations, ($n-2$)th order for the singles residual equations,  etc.}
Low-orders of UCC($n$) {are equivalent to} other familiar electronic structure approaches. For example, at $n=2$ [UCC(2)], one recovers 2nd order energy and 1st order wave function expressions from MBPT. Truncation at $n=3$ [UCC(3)] recovers the zeroth-order coupled electron pair approximation [CEPA(0), {which is} also known as linearized CC with double excitations, LCCD]. 
Truncation beyond third order generates new theories and opportunities to introduce additional approximations. For example, complete UCC(4) and UCC(5) theories should account for $\hat{T}_3$ and $\hat{T}_4$, respectively, due to these $\hat{T}_n$ contributions to the energy appearing at the 4th- and 5th-order terms in MBPT. In this work, however, we consider only singles and doubles cluster amplitudes, in which case these approaches are referred to as UCCSD($n$). 

Aside from perturbative analysis, the BCH expansion could simply be truncated according to commutator rank, {\em e.g.}, a rank-2 approach could include up to double commutators in the energy expression and single commutators in the amplitude equations, regardless of the perturbation order of the operators that appear in the commutators. We are unaware of any such approximation applied directly to Eq.~\ref{eqn:ucc_bch}, but there are several studies that apply this approach to a slightly different representation of the similarity transformation involving Bernoulli numbers.\cite{Mukherjee18_244110, Cheng21_174102, Cheng22_2281}
In these approaches, $\bar{H}$ is partitioned according to commutator rank as
\begin{align}
    \text{exp} ( -\hat{\sigma} )\hat{H} \text{exp}\left ( \hat{\sigma} \right ) = \bar{H}^0 + \bar{H}^1 + \bar{H}^2 + ...
\end{align}
with
\begin{align}
\label{EQN:H0}
        \bar{H}^0 &= f + v \\
        \label{EQN:H1}
        \bar{H}^1 &= [f, \hat{\sigma}] + \frac{1}{2} [v, \hat{\sigma}] + \frac{1}{2} [v_R, \hat{\sigma}] \\
        \label{EQN:H2}
        \bar{H}^2 &= \frac{1}{12}[[v_N, \hat{\sigma}], \hat{\sigma}] + \frac{1}{4}[[v, \hat{\sigma}]_R, \hat{\sigma}] + \frac{1}{4}[[v_R, \hat{\sigma}]_R, \hat{\sigma}] \\
        ... \nonumber
\end{align}
For a derivation of this representation, 
definitions of $\bar{H}^3$ and $\bar{H}^4$, and general recipes for constructing higher-order terms, the reader is referred to Ref.~\citenum{Mukherjee18_244110}. Compared to the usual BCH expansion, Eqs.~\ref{EQN:H1} and \ref{EQN:H2} are unique in that the operators are partitioned into pure excitation / de-excitation parts (denoted $N$ above) and the remainder of the operator (denoted $R$). Note that [de-]excitations beyond a maximum excitation order ({\em e.g.}, doubles for a UCCSD-based model) are classified as $R$-type operators in this {scheme.\cite{Thielen23_thesis}}  Note also that, in this expansion, the Fock operator does not appear in commutators of higher rank than one. As already mentioned, additional alternatives to MBPT-based truncation have been proposed, such as schemes that give the exact energy for a specific number of electrons,\cite{Bartlett06_3393} but we limit our focus here to the MBPT- and commutator-rank-based strategies.

\section{Computational Details}

\label{SEC:COMPUTATIONAL_DETAILS}

Equations and Python code corresponding to the UCCSD, CCSD, and CCSDT energy and residual equations were obtained using the p$^\dagger$q package,\cite{DePrince21_e1954709,DePrince25_arXiv_2501.08882} which is capable of generating expressions for both the standard and Bernoulli representations of the BCH expansion of the UCC $\bar{H}$, as well as standard expressions for the CC $\bar{H}$.  The autogenerated Python code was incorporated into an in-house Python-based CC/UCC solver, with all required integrals taken from the \textsc{Psi4} quantum chemistry package.\cite{Sherrill20_184108} The canonical restricted Hartree-Fock (RHF) orbitals were obtained from RHF calculations carried out using \textsc{Psi4}. Full CI calculations were also performed using \textsc{Psi4}.

Calculations  on small molecules (HF, H$_2$O, N$_2$, CO, F$_2$) were carried out using the cc-pVDZ basis set,\cite{Dunning89_1007,Wilson11_69} with equilibrium geometries taken from the NIST Chemistry Webbook.\cite{Herzberg_NIST} These geometries are compiled in the Supporting Information. In each case, calculations are carried out within the frozen-core approximation, with reference energies taken from full CI (with the exception of F$_2$, where the reference is CCSDT). We also consider all-electron calculations for the Be+H$_2$ insertion reaction, which were also carried out in the cc-pVDZ basis set, with geometries defined (in units of $a_0$) by the reaction coordinate given in Ref.~\citenum{Evangelista11_224102}, where the beryllium atom lies at the point $(0, 0, 0)$ and the hydrogen atoms lie at the points $(x, \pm y, 0)$ defined by the following equation:
\begin{equation}
\label{eqn:beh2_coordinate}
    y(x) = 2.54 - 0.46x.
\end{equation}

\section{Results and Discussion}

\label{SEC:RESULTS}

We begin by considering the quality of energies obtained from various approximations to UCCSD relative to the full CI energy for small molecules at their equilibrium geometries (1.0 $R_\text{e}$, where $R_\text{e}$ represents the equilibrium bond length) and away from equilibrium (1.5 $R_\text{e}$ and 2.0 $R_\text{e}$). Table \ref{tab:ucc_mbpt} provides energy errors for the MBPT based UCC approaches and CCSD. At equilibrium, trends across molecules are fairly consistent. For example, of the UCC approaches, UCC(3) shows the smallest error with respect to full CI, followed by UCCSD(5), UCCSD(4), and CCSD, in that order. {The only exception is for CO, where UCCSD(4) outperforms UCCSD(5) by less than 0.5 m$E_\text{h}$.} Not surprisingly, the UCC(2) energy (which is equivalent to that from 2nd order MBPT) displays the largest errors at equilibrium.

At intermediate geometries (1.5 $R_\text{e}$), UCC(2) and UCC(3) become unreliable, giving energies that are significantly below the full CI in some cases. 
{Of the remaining methods, UCCSD(4) and UCCSD(5) display comparable mean absolute errors (MAEs, $\approx 16$ m$E_\text{h}$ and $\approx 19$ m$E_\text{h}$, respectively), although w}e note difficulty in finding solutions to the UCCSD(4) amplitude equations for F$_2$ at 1.5 $R_\text{e}$ {and for UCCSD(5) in the case of CO at 1.5 $R_\text{e}$}. {Mean errors for CCSD are somewhat larger ($\approx 28$ m$E_\text{h}$)}. For some molecules (HF, H$_2$O, and F$_2$), CCSD agrees with UCCSD(5) to within 5 m$E_\text{h}$. On the other hand, the CCSD energ{y is} significantly worse than { that} from UCCSD(5) for {N$_2$
(by about 13
m$E_\text{h}$).
}

Far from equilbrium (2.0 $R_\text{e}$){,} UCC(2) and UCC(3) remain unreliable, diverging below full CI by as much as -0.291 $E_\text{h}$ [UCC(2), for N$_2$] or above full CI by as much as 1.5 $E_\text{h}$ [UCC(3), for F$_2$]. The amplitude equations for all iterative UCC methods cannot be converged for at least one system [UCC(2) is the exception because the UCC(2) amplitudes are determined non-iteratively]. Of the iterative UCC approaches, UCCSD(5) provide the greatest stability, although it has convergence issues for two systems ({HF and N$_2$}). On the other hand, we are able to converge the CCSD amplitude equations for all systems. In terms of energetics, for the cases where the amplitude equations converge, UCCSD(5) again agrees {reasonably} well with {CCSD,
with the exception of the CO molecule, where the CCSD error is about 28 m$E_\text{h}$, and the UCCSD(5) energy falls below the full CI energy by roughly 17 m$E_\text{h}$.}

\begin{table*}[!htbp]
\begin{minipage}{\textwidth}
    \centering
    \caption{
        \label{tab:ucc_mbpt}
        Errors in the electronic energies of several test molecules at their equilibrium and stretched geometries, obtained from UCC calculations with MBPT-based truncation scheme with respect to the full CI (HF, H$_2$O, N$_2$, and CO) or CCSDT (F$_2$) reference values, using the cc-pVDZ basis set. CCSD data are provided for comparison purposes. The errors are reported in units of m$E_\text{h}$, whereas the reference full CI / CCSDT data are shown in Eh.}
    \begin{tabular*}{\textwidth}{@{\extracolsep{\fill}}ccrrrrrr}
        \hline
        \hline
        System & $n \times R_\text{e}$ & UCC(2) & UCC(3) & UCCSD(4) & UCCSD(5) & CCSD & Full CI\footnotemark[1]\\
        \hline
        \multirow{3}{*}{HF}
        & 1.0 & $    7.602 $ & $    1.270 $ & $     {2.231} $ & $     {1.967} $ &  $    2.414 $ & $-100.228639$ \\
        & 1.5 & $   11.783 $ & $    2.466 $ & $     {4.337} $ & $     {3.273} $ &  $    4.665 $ & $-100.140300$ \\
        & 2.0 & $   27.404 $ & $    0.129 $ & $     {6.702} $ &  {---\footnotemark[2]} &  $   10.195 $ & $-100.063618$ \\
        \hline
        \multirow{3}{*}{H$_2$O}
        & 1.0 & $   13.240 $ & $    1.385 $ & $     {3.516} $ & $     {3.097} $ &  $    3.673 $ & $-76.241680$\\
        & 1.5 & $   22.817 $ & $    2.156 $ & $     {8.828} $ & $     {6.973} $ &  $    9.436 $ & $-76.083710$\\
        & 2.0 & $   48.748 $ & $  -18.065 $ &        {---\footnotemark[2]}    & $    {17.591} $ &  $   20.870 $ & $-75.956086$\\
        \hline
        \multirow{3}{*}{N$_2$}
        & 1.0 & $   16.557 $ & $    6.774 $ & $    {12.642} $ & $    {11.693} $ &  $   13.588 $ & $-109.276978$\\
        & 1.5 & $  -64.184 $ & $   92.824 $ & $    {39.285} $ & $    {37.819} $ &  $   50.780 $ & $-109.063009$\\
        & 2.0 & $ -290.778 $ & {55.514}   &       {343.578}    &       ---{\footnotemark[2]}    &  $   46.231 $ & $-108.968050$\\
        \hline
        \multirow{3}{*}{CO}
        & 1.0 & $   19.816 $ & $    8.313 $ & $    {10.558} $ & $    {11.006} $ &  $   12.118 $ & $-113.055853$\\
        & 1.5 & $   46.296 $ & $   59.349 $ & $   {13.031} $ &  {---\footnotemark[2]} &  $   43.555 $ & $-112.852828$\\
        & 2.0 & $   61.898 $ & $  -14.104 $ &        {---\footnotemark[2]}    & $    {-17.461} $ &  $   28.076 $ & $-112.717009$\\
        \hline
        \multirow{3}{*}{F$_2$}
        & 1.0 & $   18.166 $ & $    0.652 $                 & $     {8.951} $ & $     {7.460} $ &  $    9.294 $ & $-199.097752$ \\
        & 1.5 & $   46.827 $ & $ -214.424 $                 &        {---\footnotemark[2]}    & $    {27.514} $ &  $   32.202 $ & $-199.061585$ \\
        & 2.0 & $    1.624 $ & $ 1541.937 ${\footnotemark[3]} & $    {21.070} $ & $    {46.231} $ &  $   45.387 $ & $-199.054007$ \\
        \hline
        \multirow{3}{*}{MAE\footnotemark[4]}
        & 1.0 & 15.076 &   3.679 &   7.579 &  7.045 &  8.217 & --- \\
        & 1.5 & 38.381 &  74.244 &  16.370 & 18.895 & 28.128 & --- \\
        & 2.0 & 86.090 & 325.950 & 123.783 & 27.094 & 30.152 & --- \\

        \hline
        \hline
    \end{tabular*}
    \footnotetext[1]{For F$_2$, the reference data is CCSDT instead of full CI.}
    \footnotetext[2]{These numbers did not converge beyond 0.1 $E_\text{h}$ and are excluded from the MAE calculations.}
    \footnotetext[3]{The correlation energy is positive.}
    \footnotetext[4]{Mean absolute error.}
\end{minipage}
\end{table*}

{Before moving on, we note that some standard indicators of the reliability of CC methodologies for systems with multireference character correlate reasonably well with the behavior of the UCC($n$) approaches. Let us consider the six cases in Table \ref{tab:ucc_mbpt} where these methods do not converge. In all cases, either the T1 diagnostic\cite{Schaefer89_81} ($||t_1|| / \sqrt{n_\mathrm{o}}$, where $n_\mathrm{o}$ is the number of occupied spin orbitals) or the T2 diagnostic\cite{Piecuch07_112501} ($||t_2|| / \sqrt{n_\mathrm{o}}$) are large (at least 0.02 for the T1 diagnostic and 0.1 for the T2 diagnostic).  One or both of these diagnostics are also large for these methods in cases where the energy errors are large ({\em e.g.}, > 30 m$E_\text{h}$) or negative. These diagnostics are not perfect indicators of the performance of these methods, but they nonetheless can inform the degree to which the energies can be trusted. For example, large T2 diagnostic values may indicate the need to include higher--than--doubly-excited clusters, which are ignored in the present work.}

Table \ref{tab:ucc_bch_bernoulli} provides errors (relative to full CI) for UCCSD methods where the similarity transformation is truncated according to commutator rank; results are provided using the standard BCH expansion in Eq.~\ref{eqn:ucc_bch} and the Bernoulli expansion. In either case, the energy expression includes up to $n$ nested commutators (with $n = 2, 3, 4$), whereas the amplitude equations involve a maximum commutator rank of $n-1$. We shall refer to such approaches as rank-$n$ approximations. At equilibrium (1.0 $R_\text{e}$), rank-2 approximations universally provide the lowest errors, although it should be noted that the rank-2 energies fall below the full CI energy for two molecules {[CO (for the Bernoulli expansion) and F$_2$ (for both the Bernoulli and BCH expansions)]}. It is also noteworthy that the rank-2 approximations built upon the standard and Bernoulli-based expansions of $\bar{H}$ agree with one another to less than {0.06 m$E_\text{h}$, except in the case of the CO molecule where the errors also have opposite signs}. In terms of energy errors, the next-best-performing methods{, within each $\bar{H}$ expansion scheme,} are the rank-4 and rank-3 approximations, in that order. {Note that, for CO, the Bernoulli-based rank-3 scheme slightly outperforms the standard BCH-based rank-4 scheme. Nonetheless}, for a given maximum commutator rank, the approximations built upon the standard and Bernoulli-based expansions of $\bar{H}$ agree well with one-another; the mean absolute difference between the two forms of the rank-3 and rank-4 approximations are {2.425} m$E_\text{h}$ and {0.842} m$E_\text{h}$, respectively. 

At intermediate geometries (1.5 $R_\text{e}$), the rank-2 approaches begin to fail, with energy errors that exceed $-300$ m$E_\text{h}$ in the case of F$_2$. These failures are not surprising, given the similar failures of UCC(2) in Table \ref{tab:ucc_mbpt}, although the energy errors are significantly larger for the rank-2 approaches. Note that the UCC(2) amplitude equations are solved non-iteratively, whereas the rank-2 amplitude equations must be solved iteratively, due to the presence of $\hat{\sigma}_1$ and $\hat{\sigma}_2$ in the doubles and singles residual equations, respectively. As for the rank-3 and rank-4 approaches, {we observe similar trends where the rank-4 approaches outperform the rank-3 ones. However, Bernoulli-based rank-3 scheme provides significantly lower energy errors than the standard BCH-based rank-3 scheme, with the quality of the energy approaching that of the standard BCH-based rank-4 scheme in some cases (CO and F$_2$).} For the Bernoulli expansion, the rank-4 approximation consistently outperforms the rank-3 approximation, by as much as 14 m$E_\text{h}$ in the case of N$_2$.
{The situation is similar for the standard BCH expansion, where the improvements brought on by rank-4 approaches can be as large as 27 m$E_\text{h}$ in the case of N$_2$.}
The rank-4 approximation to the standard BCH expansion behaves similarly to that for the Bernoulli expansion, with the largest difference in energies approaching {5} m$E_\text{h}$ for CO and {F$_2$.}

Far from equilibrium (2.0 $R_\text{e}$), the rank-2 approximations continue to be unreliable; as an example, energy errors exceed 1 $E_\text{h}$ for F$_2$ using both $\bar{H}$ expansions.
{ The performance of both $\bar{H}$ expansion variants are similar, with the exception of CO, where the difference between the rank-2 approaches amounts to more than 30 m$E_\text{h}$.}
{W}e observe no stability issues in any of the rank-$n$ approaches, regardless of the representation of the BCH expansion.  For the Bernoulli expansion, the rank-4 approximation outperforms the rank-3 approximation in most cases (by up to $\approx$ 11 m$E_\text{h}$ for CO), with the exception of H$_2$O, where the rank-3 approximation has a smaller error by about 1 m$E_\text{h}$. We find that the rank-4 approximation to the standard BCH expansion gives similar and slightly better energetics than the rank-4 approximation to the Bernoulli expansion for HF and H$_2$O. The remaining cases are a toss up; the standard BCH expansion leads to significantly better energetics in the cases { of
CO}, while the Bernoulli expansion clearly outperforms the standard expansion for { N$_2$ and} F$_2$. 
{As for the other geometries, at 2.0 $R_\text{e}$, the rank-4 approaches generally outperform the rank-3 ones, within a given $\bar{H}$ expansion scheme. The rank-4 approaches, with MAEs of 58--63 m$E_\text{h}$, provide better agreement with full CI, as compared to the corresponding rank-3 methods that have MAEs of 67--71 m$E_\text{h}$. Furthermore, the rank-4 approaches reproduce the CCSD energetics to within less than 1 m$E_\text{h}$ (HF and H$_2$O) or a few m$E_\text{h}$ (F$_2$), with the exception of CO and N$_2$ where the deviations are as large as 0.05--0.1 $E_\text{h}$.}

Comparing the commutator-rank-based and MBPT-based schemes, we can make the following observations based on the data in Tables \ref{tab:ucc_mbpt} and \ref{tab:ucc_bch_bernoulli}. First, in general, UCC(2), UCC(3), and the rank-2 approximations provide reasonable results at equilibrium but quickly become unreliable at non-equilibrium geometries. For this reason, the remainder of this discussion will focus on the higher-order approximations. Second, the MBPT-based truncation scheme is far more prone to numerical stability issues. We find many cases at stretched geometries where 4th and 5th order UCC approaches fail to converge. 
On the other hand, no such issues plague the rank-3 and rank-4 approaches, at least for the systems in Table \ref{tab:ucc_bch_bernoulli}. Third, given the stability issues in the MBPT-based approaches, it appears that rank-4 approximations more readily reproduce the results of CCSD, although there are non-negligible differences between the respective energetics far from equilibrium, as discussed above. {Based on the data at 1.5 $R_\text{e}$ and 2.0 $R_\text{e}$, we can also see that the Bernoulli-based rank-3 scheme generally outperforms the corresponding standard BCH-based scheme, which suggests that the Bernoulli expansion may offer slightly more rapid convergence toward the UCCSD limit, with increasing commutator rank.}

\begin{table*}[!htbp]
    \centering
    \caption{
        \label{tab:ucc_bch_bernoulli}
        Errors in the electronic energies of several test molecules at their equilibrium and stretched geometries, obtained from different UCC calculations with standard and Bernoulli-based BCH truncation schemes with respect to the full CI (HF, H$_2$O, N$_2$, and CO) or CCSDT (F$_2$) reference values in Table \ref{tab:ucc_mbpt}, using the cc-pVDZ basis set. The errors are reported in units of m$E_\text{h}$.}
    \begin{tabular*}{\textwidth}{@{\extracolsep{\fill}}ccrrrrrr}
        \hline
        \hline
        \multirow{3}{*}{System} & \multirow{3}{*}{$n \times R_\text{e}$}  & \multicolumn{6}{c}{Energy Expression Maximum Commutator Rank} \\
        \cline{3-8}
        & & \multicolumn{3}{c}{Standard} & \multicolumn{3}{c}{Bernoulli} \\
        \cline{3-5} \cline{6-8}
        & & 2 & 3 & 4 & 2 & 3 & 4\\
        \hline
        \multirow{3}{*}{HF} 
        & 1.0 & {  0.513} & { 3.054} & {1.973} &   0.493 &  2.620 & 2.070 \\
        & 1.5 & { -1.944} & { 6.582} & {3.974} &  -2.165 &  5.235 & 4.243 \\
        & 2.0 & {-28.453} & {16.343} & {9.251} & -31.928 & 10.824 & 9.698 \\  
        \hline
        \multirow{3}{*}{H$_2$O}
        & 1.0 & {  0.456} & { 4.698} & { 3.114} &   0.446 &  3.923 &  3.248 \\
        & 1.5 & { -5.794} & {13.152} & { 8.038} &  -5.870 & 10.472 &  8.523 \\
        & 2.0 & {-49.722} & {33.488} & {20.277} & -50.879 & 19.835 & 20.750 \\
        \hline
        \multirow{3}{*}{N$_2$}
        & 1.0 & { 2.337} & {16.936} & {11.796} &  2.277 & 14.524 & 12.039 \\
        & 1.5 & {76.905} & {64.918} & {38.300} & 76.395 & 52.890 & 38.785 \\
        & 2.0 & {-0.349} & {58.118} & {91.098} & -3.138 & 91.962 & 85.780 \\
        \hline
        \multirow{3}{*}{CO}
        & 1.0 & {  2.048} & { 15.296} & { 11.251} &  -1.864 &   9.639 &   7.826 \\
        & 1.5 & {169.072} & { 57.847} & { 46.905} & 211.102 &  48.607 &  42.348 \\
        & 2.0 & {142.343} & {177.987} & {114.241} & 139.207 & 169.185 & 158.017 \\
        \hline
        \multirow{3}{*}{F$_2$}
        & 1.0 & {  -5.683} & {12.982} & { 7.801} &   -5.637 & 10.134 &  8.111 \\
        & 1.5 & {-365.770} & {48.075} & {32.248} & -364.694 & 33.682 & 30.288 \\
        & 2.0 & {1228.723} & {69.355} & {53.961} & 1227.811 & 44.927 & 42.626 \\
        \hline
        \multirow{3}{*}{MAE\footnotemark[1]}
        & 1.0 &   2.207 & 10.593 &  7.187 &   2.144 &  8.168 &  6.659 \\
        & 1.5 & 123.897 & 38.115 & 25.893 & 132.045 & 30.177 & 24.837 \\
        & 2.0 & 289.918 & 71.058 & 57.766 & 284.513 & 67.347 & 63.374 \\
        \hline
        \hline
    \end{tabular*}
    \footnotetext[1]{Mean absolute error.}
\end{table*}

\begin{figure}[!htpb]
    \centering
    \includegraphics[width=\linewidth]{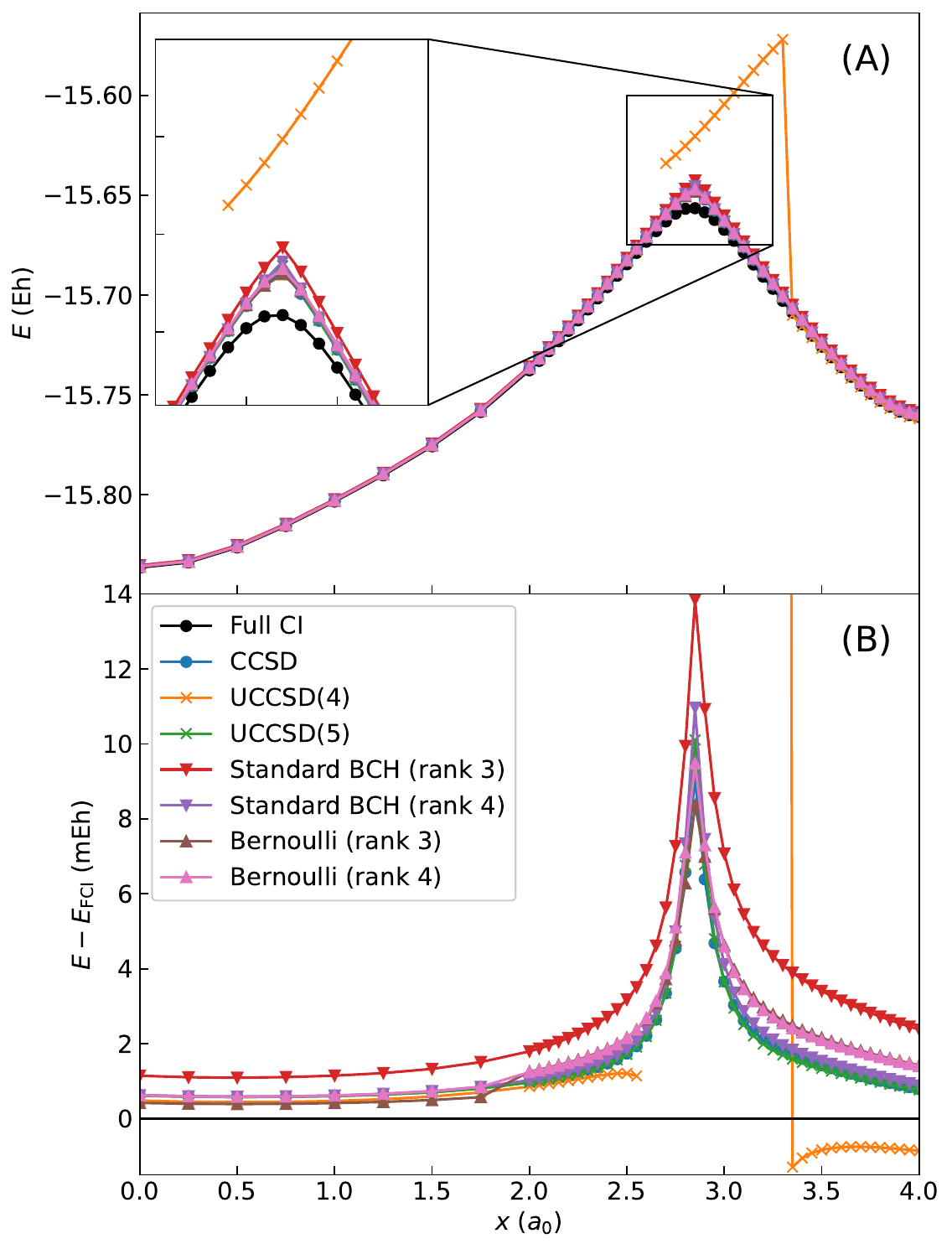}
    \caption{
        \label{fig:beh2_ucc}
        The PECs of BeH$_2$ insertion model obtained using full CI, CCSD, and the UCC approaches examined in this work (A), along with the corresponding errors of the PECs relative to the full CI reference (B).
    }
\end{figure}

We now consider the C$_{2v}$ insertion of Be into H$_2$, which has long served as a benchmark problem for multireference electronic structure methods.\cite{Mukherjee98_163,Mukherjee99_6171,Surjan02_980,Ahlrichs88_413,Gellene00_10951,Todorov05_2497,Carsky04_211,Adamowicz06_427} Figure \ref{fig:beh2_ucc}(A) depicts computed potential energy curves (PECs) for this reaction, along the reaction coordinate taken from Ref.~\citenum{Evangelista11_224102} (which is reproduced in Sec.~\ref{SEC:COMPUTATIONAL_DETAILS}). Energy errors with respect to full CI are illustrated in Fig.~\ref{fig:beh2_ucc}(B) and tabulated in the Supporting Information. Outside of the multireference part of the PEC ({\em i.e.}, $x < 2.5$ $a_0$ and $x > 3.5$ $a_0$, CCSD and the approximations to UCCSD agree well with the full CI, giving errors that are at most $\approx$ 2 m$E_\text{h}$ in magnitude. While the magnitudes of the errors are small, UCCSD(4)
{ yields energies that are below those from full CI at large $x$. In the same region, the rank-3 approximation to UCCSD that uses the standard BCH expansion overestimates full CI energy with errors that are roughly twice as large as the other UCCSD approaches shown in Fig.~\ref{fig:beh2_ucc}(B).}
In the multireference region, { UCCSD(4) begins} to fail altogether. {Indeed, 
we} are unable to converge the amplitude equations for UCCSD(4) in the region $x = $ 2.60 $a_0$ -- {2.65} $a_0$. In addition, from $x =$ {2.70  $a_0$ --  3.30  $a_0$}, the UCCSD(4) amplitude equations converge to a high-energy solution, even when the calculations are seeded by converged amplitudes from {CCSD.
Furthermore, judging by the pattern in errors of UCCSD(4) relative to full CI, the UCCSD(4) PEC seems to start diverging toward $-\infty$ in the MR region, which is consistent with our earlier observations that this UCC approach is not numerically stable.
The} rank-4 UCCSD approach with the standard BCH expansion { works well}, 
resulting in a PEC that is similar in quality to that obtained from CCSD or UCCSD(5){, whereas its rank-3 approximation produces a somewhat parallel curve that are a few m$E_\text{h}$ higher in energy}. Interestingly, the rank-3 and rank-4 Bernoulli BCH approaches are both numerically stable in the multireference region and provide comparable results, which is consistent with our previous observation that the Bernoulli form of the BCH expansion {may represent the more rapidly} convergent commutator-rank-based truncation scheme. Lastly, while the CCSD, UCCSD(5),
{ rank-3 and rank-4 standard BCH, and rank-3 and rank-4 Bernoulli expansion}
approaches are all numerically well-behaved in the multireference region, these methods nonetheless all present the derivative discontinuity in the PEC at $x = $ 2.85 $a_0$ that we expect from a single-reference electronic structure methods. At this point, these methods display their maximum errors with respect to full CI ({9.096 m$E_\text{h}$, 10.106 m$E_\text{h}$, 13.854 m$E_\text{h}$, 10.966} m$E_\text{h}$, 8.365 m$E_\text{h}$, {and} 9.479 m$E_\text{h}$, respectively).

\section{Conclusions}

\label{SEC:CONCLUSIONS}

The UCC theory represents one of several attempts to realize a useful Hermitian alternative to conventional CC theory. However, the non-truncating nature of the BCH expansion in UCC has precluded the widespread adoption of the approach and has led to the development of multiple strategies for realizing compact and reliable schemes for truncating the UCC $\bar{H}$. In this work, we have directly compared the numerical properties of approximate UCCSD approaches built upon truncation schemes that rely on MBPT- and commutator-rank-based analyses.

In general, we find that low-order MBPT-based approaches [UCC(2) and UCC(3)] and the rank-2 approaches can provide reasonable results at equilibrium geometries, but they quickly become unreliable away from equilibrium. Moving to higher orders in MBPT does not always improve the situation, as the solutions to the amplitude equations for UCCSD(4) and UCCSD(5) cannot always be found. For the cases where UCCSD(5) solutions do exist, though, this approach does seem to provide reasonable agreement with CCSD in many cases. On the other hand, the higher-order commutator-rank-based approaches are more robust in the sense that they do not display any of the convergence issues we have observed for UCC($n$), at least for the cases studied in this work. 

When comparing rank-3 and rank-4 approaches based on standard or Bernoulli representations of the BCH expansion, {there is evidence} that the Bernoulli representation offers the more rapid convergence to the UCCSD limit{. The data in Table \ref{tab:ucc_bch_bernoulli} show that energies from the rank-3 and rank-4 Bernoulli expansion schemes more closely resemble one another, as compared to the energies from the rank-3 and rank-4 standard BCH expansions. The similarities between the Bernoulli-based rank-3 and rank-4 schemes are even more apparent in the energy errors for the Be + H$_2$ in Fig.~\ref{fig:beh2_ucc}.} 
Put together, the data presented in this work suggest that the rank-3 and rank-4 schemes that use the Bernoulli representation of the BCH expansion {and the rank-4 scheme that employs the standard BCH expansion} are the most robust and reliable approximate UCCSD approaches. {The Bernoulli-based rank-3 approach provides better energetics than the rank-3 scheme that uses the standard BCH expansion, so we conclude that the former is the more reliable low-rank scheme. } The {Bernoulli-based} rank-3 approach is equivalent to the quadratic UCCSD (qUCCSD) approach employed in Refs.~\citenum{Cheng21_174102} and \citenum{Cheng22_2281}, so this work also offers numerical justification for the truncation scheme chosen in those works. 

{In terms of computational cost, all of the approximate UCCSD approaches considered in this work share the $\mathcal{O}(N^6)$ scaling exhibited by CCSD (where $N$ is a measure of system size). Using automated software generation tools (the p$^\dagger$q package\cite{DePrince25_arXiv_2501.08882}) to analyze the contractions that scale as $\mathcal{O}(n_\mathrm{o}^2n_\mathrm{v}^4)$ and $\mathcal{O}(n_\mathrm{o}^3n_\mathrm{v}^3)$, where $n_\mathrm{v}$ refer to the number of orbitals that are empty in the reference configuration, we find that the rank-3 approaches and UCCSD(4) carry roughly twice the floating-point cost of CCSD (this analysis is consistent with that in Ref.~\citenum{Cheng22_2281} for qUCCSD). Our analysis also indicates that the rank-4 approaches and UCCSD(5) exhibit a floating-point cost that is roughly an order of magnitude greater than that for CCSD. This estimate is probably an upper bound, as the equations generated by p$^\dagger$q likely do not correspond to the ideal implementations of these methods. Regardless, when taking computational efficiency into consideration, it appears that the marginal improvement afforded by the rank-4 Bernoulli approach is not worth the additional cost compared to the rank-3 version / qUCCSD. } 

\vspace{0.5cm}

{\bf Supporting Information} The equilibrium geometries for the molecules examined in this paper, and energy errors for CCSD and approximate UCCSD methods relative to full CI for the Be+H$_2$ insertion reaction.

\vspace{0.5cm}

\begin{acknowledgments} The authors would like to congratulate Professor Piotr Piecuch for his 65th birthday and for the well-deserved festschrift celebrating his contribution to the field of quantum chemistry, especially in the development of CC methodologies. SHY is deeply grateful for the mentoring from and friendship with Professor Piotr Piecuch during his graduate school years and beyond. This material is based upon work supported by the National Science Foundation under Grant No. OAC-2103705.\\ 
\end{acknowledgments}

\bibliography{main}

\end{document}